# $^{26}$Al and $^{60}$Fe from Supernova Explosions

F. X. Timmes[1,2,3,4], S. E. Woosley[2,3], D. H.Hartmann[4],
R. D. Hoffman[2], T. A. Weaver[3], & F. Matteucci[5]

[1] University of Chicago, LASR, Chicago IL. 60637, USA
[2] UC Santa Cruz, Astronomy Department, Santa Cruz, CA 95064, USA
[3] Lawrence Livermore National Laboratory, General Studies Division, Livermore CA. 94550, USA
[4] Clemson University, Dept. of Physics and Astronomy, Clemson, SC. 29634, USA
[5] European Southern Observatory, K. Schwartzchild. Str. 2, D-8046, Garching, Germany

**Abstract**

Using recently calculated yields for Type II supernovae, along with models for chemical evolution and the distribution of mass in the interstellar medium, the current abundances and spatial distributions of two key gamma-ray radioactivities, $^{26}$Al and $^{60}$Fe, are determined. The estimated steady state production rates are $2.0 \pm 1.0$ M$_\odot$ Myr$^{-1}$ for $^{26}$Al and $0.75 \pm 0.4$ M$_\odot$ Myr$^{-1}$ for $^{60}$Fe. This corresponds to $2.2 \pm 1.1$ M$_\odot$ of $^{26}$Al and $1.7 \pm 0.9$ M$_\odot$ of $^{60}$Fe in the present interstellar medium. Sources of uncertainty are discussed, one of the more important being the current rate of core collapse supernovae in the Galaxy. Our simple model gives three per century, but reasonable changes in the star formation rate could easily accommodate a core collapse rate one-half as large, and thus one-half the yields. When these stellar and chemical evolution results are mapped into a three dimensional model of the Galaxy, the calculated 1809 keV gamma-ray flux map is consistent with the *Compton Gamma Ray Observatory* observations of a steep decline in the flux outside a longitude of $\pm$ 50° from the Galactic center, and the slight flux enhancements observed in the vicinity of spiral arms. Other potential stellar sources of $^{26}$Al and $^{60}$Fe are mentioned, especially the possibility of $^{60}$Fe synthesis in Type Ia supernovae. Predictions for the $^{60}$Fe mass distribution, total mass, and flux map are given.

1. Introduction

Because their lifetimes are long compared to the interval between supernovae, yet short enough to give rise to detectable emission when they decay, the species $^{26}$Al ($\tau_{1/2} = 7.5 \times 10^5$ yr; Tuli 1990) and $^{60}$Fe ($\tau_{1/2} = 1.5 \times 10^6$ yr; Tuli 1990) have long been prime candidates for gamma-ray astronomy (Clayton 1971, 1973, 1982, 1983; Ramaty & Lingenfelter 1977; Woosley & Weaver 1980; Gehrels *et al.* 1993). Further, since the decay time is short compared to Galactic rotation, the abundances of these species in the interstellar medium serves as an important tracer of the stellar population responsible (e.g., massive stars) for their synthesis (e.g., Prantzos 1991, 1993a,b).

Gamma-ray photons from $^{26}$Al remain the only radionuclide to have been detected, at a $3\sigma$ level, in the interstellar medium (Mahoney *et al.* 1982, 1984; Share *et al.* 1985; Gehrels *et al.* 1993; Diehl *et al.* 1993). The observed 1809 keV $\gamma$-ray emission is produced principally by $\beta^+$ decay (branching ratio 82%) and secondarily from electron capture (branching ratio 15%) from the $J^\pi = 5^+$ ground state of $^{26}$Al to the first excited level ($J^\pi = 2^+$) level of $^{26}$Mg, which then de-excites (at the same $\gamma$-energy) to the $J^\pi = 0^+$ ground state of $^{26}$Mg.

Enhanced $^{26}$Mg/$^{24}$Mg ratios in the Ca-Al rich inclusions of the Allende meteorite was the first evidence for live $^{26}$Al in the early solar system (Lee, Papanastassiou, & Wasserburg 1977). Extensive measurements of other meteorites (Anders & Zinner 1993; Ott 1993) have confirmed the anomalous magnesium isotopic ratio in addition to discovering many other isotopic anomalies. Observations of the diffuse emission in the interstellar medium from 1809 keV flux maps suggest an abundance of $\sim 2$ M$_\odot$ of live $^{26}$Al in the interstellar medium of our Galaxy (Mahoney *et al.* 1982, 1984; Share *et al.* 1985; Gehrels *et al.* 1993; Diehl *et al.* 1993). This estimate may be reduced somewhat if a major portion of the signal comes from nearby sources (Chen, Gehrels & Diehl 1995; Hartmann *et al.* 1995). Gamma-rays from $^{60}$Fe have not yet been detected, although searches at moderate sensitivity have been carried out (Leising & Share 1994).

It is presently controversial whether the observed $^{26}$Al is chiefly the product of supernovae in massive stars, or whether other components such as novae (Clayton & Hoyle 1976; Clayton 1984; Weiss & Truran 1990; Nofar, Shaviv, & Starrfield 1991; Starrfield *et al.* 1993), asymptotic giant branch stars (Norgaard 1980; Cameron 1984, 1993; Clayton 1985; Clayton & Leising 1987; Forestini, Paulus, & Arnould 1991; Bazan *et al.* 1993; Wasserburg *et al.* 1994), or the winds of Wolf-Rayet stars (Dearborn & Blake 1984; Prantzos & Casse 1986; Walter & Maeder 1987), might be required (technically this last contribution should be counted along with the $^{26}$Al made by the same star when it explodes).

The main result of this paper will be to show that massive stars, specifically the explosion of Type II supernovae in the 11 – 40 M$_\odot$ range, *can* produce adequate amounts of $^{26}$Al, with a spatial distribution that is consistent with all the observations. This result is derived from a stellar-chemical evolution model that gives reasonably consistent nucleosynthesis for all the stable isotopes from hydrogen to zinc (Timmes, Woosley, & Weaver 1995). This result does not necessarily imply that all $^{26}$Al is produced this way. There are large uncertainties both in our results and those of others. However, having achieved this result without artificially tuning either the stellar nucleosynthesis or chemical evolution models, we are encouraged to make predictions regarding the abundance and spatial distribution of $^{60}$Fe, which is another long-lived isotope made, for the most part, by the same stars. The total mass of live $^{60}$Fe in the Galaxy that we predict may be slightly below the sensitivity threshold of gamma-ray instruments presently in space, but the next generation of detectors should find it.

## 2. Nucleosynthesis

We briefly summarize the production of $^{26}$Al and $^{60}$Fe in massive stars and supernovae using the results of a recent 25 $M_\odot$ model as an example. The yields used in the calculation of the Galactic $^{26}$Al and $^{60}$Fe production rates come from a much larger grid of models of varying mass and metallicity calculated by Woosley & Weaver (1995). The most relevant models are those of the current epoch, i.e., solar metallicity stars. The yields of $^{26}$Al and $^{60}$Fe are summarized in Table 1. The exploded 25 $M_\odot$ model discussed below had a final kinetic energy at infinity of $1.2 \times 10^{51}$ ergs, and left a bound remnant of 2.07 $M_\odot$ (baryonic mass after fall back). The $^{56}$Ni mass ejected was 0.13 $M_\odot$.

### 2.1  $^{26}$Al Production in Massive Stars

The nucleus $^{26}$Al is produced by proton capture on $^{25}$Mg, and destroyed by $e^+$ decay, (n,p), (n,$\alpha$), and (p,$\gamma$) reactions (the outcome being sensitive to the free nucleon abundances and temperature). In a typical solar metallicity 25 $M_\odot$ model, Figure 1 shows that substantial production of $^{26}$Al occurs both before and during the explosion. Presupernova synthesis occurs in the hydrogen burning shell. Part of this $^{26}$Al is transported into the extended red-giant envelope by convective dredge up, and part remains behind in the helium core where it slowly decays. A larger amount of $^{26}$Al is made in the oxygen and neon burning shells (these frequently combine into a single burning shell during the late stages of stellar evolution; Woosley & Weaver 1980; Weaver & Woosley 1995). In these (pre-explosive) neon-oxygen burning shells there is a competition between $^{26}$Al production by proton capture on $^{25}$Mg and destruction by neutron capture on $^{26}$Al (with protons and neutrons principally provided by $^{23}$Na($\alpha$,p)$^{26}$Mg and $^{22}$Ne($\alpha$,n)$^{26}$Mg, respectively).

The $^{26}$Al abundance present in the hydrogen shell is ejected unmodified in the explosion or, perhaps in more massive stars, by a stellar wind. The explosive yield in the oxygen-neon shell is further enhanced by $\sim 30\%$ due to operation of the $\nu$-process (Woosley *et al.* 1990; Timmes *et al.* 1995; Woosley & Weaver 1995). Protons liberated by $\nu$-spallation of the abundant pre-explosive isotopes ($^{20}$Ne, $^{16}$O, $^{23}$Na, $^{24}$Mg) capture on $^{25}$Mg to produce $^{26}$Al. In stars of different mass, particularly those in which the burning shells are located nearer to the collapsing core (and thus experience a higher $\nu$-flux), this enhancement may be as high as $\sim 50\%$. The importance of the $\nu$-process contribution depends on the peak $\mu$ and $\tau$ neutrino temperature, which while uncertain, is not a free parameter. Some have suggested that the effective $\mu$ and $\tau$ neutrino temperature should be substantially less than 8 MeV employed by Woosley & Weaver (1995), perhaps more like 6 MeV (Janka & Hillebrandt 1989; Janka 1991). The problem is that the actual temperature distribution is not a blackbody distribution at any temperature. An additional difficulty is that few neutrino transport calculations have been carried to sufficiently late time (at least 3 s) or have sufficient neutrino energy resolution to see the hardening of the spectrum that occurs as the proto-neutron star cools (Janka 1991; Wilson & Mayle 1993). Recent calculations by Wilson reported in Woosley *et al.* (1994) suggest that a temperature for the $\mu$ and $\tau$-neutrinos of 8 MeV is not unrealistic (the numbers in their Fig. 3 should be divided by $\sim 4$ since the average $<\epsilon_\nu^2>/<\epsilon_\nu>$ is plotted).

In the hydrogen and especially in the oxygen-neon shell burning regions, the convective coupling between mass zones is important – convection can simultaneously bring light reactants and seed nuclei into the hot zone to aid in the synthesis, and then remove the fragile product from the high temperature region where it might otherwise be destroyed. This is why the $^{26}$Al abundance throughout the oxygen-neon shell is not flat in Figure 1, but declines starting at the base of the oxygen-neon shell.

The convective burning that takes place in the oxygen shell of a 20 M$_\odot$ star prior to core collapse was examined in two dimensions by Bazan & Arnett (1994). They find that plume structures dominate the velocity field, and that significant mixing beyond the boundaries defined by mixing-length theory (i.e., "convective overshoot") brings fresh fuel (carbon) into the convective region causing local hot spots of nuclear burning. This general picture is dramatically different from the situation encountered in one-dimensional computations of stellar evolution. The chemical inhomogeneities and local burning are likely to change the quantitative yields of several isotopes from a single star. However, any nonmonotonic and/or stochastic nature of the nucleosynthetic yields as a function of stellar mass tend to be smoothed out by integration over an initial mass function. Thus, the general features of the integrated yields, as determined from one dimensional models, will probably remain intact.

## 2.2 $^{60}$Fe Production in Massive Stars

In the solar metallicity 25 M$_\odot$ presupernova star, $^{60}$Fe is produced explosively in the oxygen-neon burning shell (see Fig. 2), and to a lesser extent in a thin layer at the base of the He-burning shell. In both cases, neutron capture on initial $^{56,58}$Fe during s-processing in helium shell burning is responsible for the presupernova abundance. The supernova produces (by the same neutron capture pathway) almost equal amounts of $^{60}$Fe in explosive helium burning and at the base of the high-temperature oxygen-neon burning shell, (in the 25 M$_\odot$ star the values were 1.18 × 10$^{-5}$ M$_\odot$ exterior to 6 M$_\odot$ and 9.3 × 10$^{-6}$ M$_\odot$ interior to 6 M$_\odot$.) The operation of the $\nu$-process did not contribute directly to the abundance of this isotope.

It is important to recognize that since $^{60}$Fe and $^{26}$Al are coproduced in the same regions within Type II supernovae, as shown by Figures 1 and 2. These two isotopes should then have similar spatial distributions in the supernova ejecta, which has important consequences for $\gamma$-line flux observations of these radioactive isotopes.

There are several interesting parallels between $^{60}$Fe and $^{26}$Al, such as the indication that live $^{60}$Fe also existed in the early solar system. Shukolyukov & Lugmair (1992, 1993) found evidence for live $^{60}$Fe in the Chernvony Kut meteorite. The measured excess of $^{60}$Ni, after alternative modes of production such as spallation and (n,$\gamma$) reactions on $^{59}$Co could be eliminated, leads to a $^{60}$Fe/$^{56}$Fe ratio at the time of iron-nickel fractionation of $\sim$ 7.5 × 10$^{-9}$. This is consistent with the inferred 10 million year hiatus before the formation of the Ca-Al rich inclusions in the Allende meteorite, which has a much larger $^{60}$Fe/$^{56}$Fe ratio at the time of fractionation of $\sim$ 1.6 × 10$^{-6}$ (Birck & Lugmair 1988). However, the possibility that some of the $^{60}$Ni in the Ca-Al rich inclusions is fossil rather than from the in situ decay of $^{60}$Fe cannot be excluded.

The yields of $^{60}$Fe and $^{26}$Al as a function of main sequence progenitor mass are shown in Figure 3. An order of magnitude estimate for the injection rate is simply the Galactic Type II supernovae rate times the mass ejected by each Type II supernovae. From Figure 3 one has M($^{26}$Al) $\approx$ 1 × 10$^{-4}$ M$_\odot$ and M($^{60}$Fe) $\approx$ 4 × 10$^{-5}$ M$_\odot$. Adopting a rate of 3 core collapse events per century, one has $\dot{\rm M}$($^{26}$Al) $\approx$ 3.0 M$_\odot$ Myr$^{-1}$ and $\dot{\rm M}$($^{60}$Fe) $\approx$ 1.2 M$_\odot$ Myr$^{-1}$. In the next section this simple order of magnitude estimate is refined, checking that no other isotope is grossly under- or overproduced.

## 3. Galactic Chemical Evolution

The $^{26}$Al and $^{60}$Fe yields, along with those of all the stable isotopes lighter than zinc, have been incorporated into a Galactic chemical evolution code (Timmes *et al.* 1995). The models used to represent the dynamical and isotopic evolution are simple and relatively standard. Essentially, each radial zone in an exponential disk begins with zero gas and accretes primordial or near-primordial material over a 2–4 Gyr *e*-folding timescale. The isotopic evolution at each radial coordinate is calculated using "zone" models (as opposed to hydrodynamic models) of chemical evolution. Standard auxiliary quantities such as a Salpeter (1955) initial mass function and a Schmidt (1959, 1963) birth-rate function were used. The model employs abundance yields for Type Ia supernovae (Nomoto, Thielemann, & Yokoi 1984; Thielemann, Nomoto, & Yokoi 1986), intermediate-low mass stars (Renzini & Voli 1981) as well as the Type II supernova of Woosley & Weaver (1995) discussed earlier. The combined stellar-chemical evolution model is in excellent agreement with the Anders & Grevesse (1989) solar abundances and is characterized by a current Galactic Type II + Ib supernova rate of 3 per century and a Type Ia rate of 0.5 per century. These calculated supernova rates, which are not an input parameter, are in good agreement with the estimates of van den Bergh & Tammann (1991) and van den Bergh & McClure (1994).

Figure 4 shows the present epoch production rate of $^{26}$Al and $^{60}$Fe (both times $2\pi r$) as a function of Galactic radius. The shape of these injection rate curves is a direct reflection of the assumed exponential Galactic disk; they should not be interpreted as *a priori* determinations of the radial distributions. There is also the implicit assumption being made that the $^{26}$Al does not move very far away from its nucleosynthetic origin site during its mean lifetime. The figure shows that the distribution of these radionuclides will be concentrated toward the center of the Galaxy. Integration of the curves in Figure 4 yield the total production rates. Including the uncertainties (discussed below), the integration gives present epoch injection rates of $2.0 \pm 1.0$ M$_\odot$ Myr$^{-1}$ for $^{26}$Al and $0.75 \pm 0.4$ M$_\odot$ Myr$^{-1}$ for $^{60}$Fe. The absolute abundance of either isotope is given, in steady state, by multiplying the injection rate by the respective mean lifetime – one then has a total of $2.2 \pm 1.1$ M$_\odot$ of $^{26}$Al and $1.7 \pm 0.9$ M$_\odot$ of live $^{60}$Fe in the present interstellar medium. Chemical evolution calculations showed that after 10–20 billion years of Galactic evolution the steady state assumption for $^{26}$Al and $^{60}$Fe is a very good approximation (differences of $\sim 1$ part in $10^2$).

Accretion of primordial or near-primordial material is included in the chemical evolution model. Growth of the Galactic disk by infall dilutes the abundance of stable $^{27}$Al, but does not directly effect the abundance of the unstable $^{26}$Al. This accounts for simultaneously calculating the correct solar abundance of $^{27}$Al (Timmes *et al.* 1995) and the present-day $^{26}$Al/$^{27}$Al ratio (Timmes, Woosley, & Weaver 1993; Clayton, Hartmann, & Leising 1993).

The steady state flux ratio of $^{60}$Fe to $^{26}$Al is given by

$$R = \frac{\dot{\mathrm{M}}(^{60}\mathrm{Fe})}{\dot{\mathrm{M}}(^{26}\mathrm{Al})} \times \frac{26}{60} \ . \tag{1}$$

Using the injection rates given above, the flux ratio is calculated to be 0.16. Since $^{60}$Fe and $^{26}$Al are coproduced in Type II + Ib events, they should have the same spatial distribution in the Galaxy. At these flux levels, $^{60}$Fe might be detectable by *Compton Gamma Ray Observatory* and should easily be detectable by INTEGRAL. Leising & Share (1994) searched almost 10 yrs of data from the *Solar Maximum Mission* Gamma-Ray Spectrometer for evidence of $\gamma$-ray line emission from the decay of the shorter-lived daughter $^{60}$Co of nucleosynthetic $^{60}$Fe. They found no direct evidence of emission, which formed their upper limit of 1.7 M$_\odot$ of $^{60}$Fe in the interstellar medium today. This upper limit is close to what our stellar-chemical evolution calculations yield. Their preferred value of 0.9 M$_\odot$ of $^{60}$Fe is based on other nucleosynthesis arguments, but within the uncertainties of our estimate.

The calculated $^{26}$Al and $^{60}$Fe injection rates and flux ratio are dominated by the input Type II nucleosynthesis. The W7 Type Ia supernovae model of Nomoto, Thielemann & Yokoi (1984) primarily ejects radioactive $^{56}$Ni and $^{56}$Co, but it also produces $3.8 \times 10^{-6}$ M$_\odot$ of $^{26}$Al and $2.3 \times 10^{-9}$ M$_\odot$ of $^{60}$Fe. Since these values are much smaller than the massive star yields (see Table 1), standard carbon deflagration models for Type Ia supernovae do not significantly contribute to the calculated abundances or injection rates. Contributions from the W7 Type Ia model were included in the calculation, however.

Recent calculations of Type Ia nucleosynthesis by Woosley & Eastman (1995), suggest that slow initial propagation of the flame in carbon-oxygen white dwarfs igniting near the Chandrasekhar mass will produce a substantial amount of $^{60}$Fe. This synthesis, typically 0.002 – 0.006 M$_\odot$, occurs in the inner few hundredths of a solar mass of the white dwarf (which is completely disrupted) where electron capture leads to an electron mole number $Y_e \approx 0.42$. This same region is also responsible for producing a number of other neutron-rich species – $^{48}$Ca, $^{50}$Ti, $^{54}$Cr, and some $^{66}$Zn – which are not made in adequate quantities elsewhere. Of all Type Ia supernova events, how many would explode from a high ignition density is very uncertain, though all current models that ignite carbon in excess of $3 \times 10^9$ g cm$^{-3}$ would qualify. Further questions such as – What fraction of Type Ia supernovae are sub-Chandrasekhar mass (e.g., Livne & Glasner 1991; Woosley & Weaver 1994), and is there any initial laminar flame propagation mode in white dwarfs that ignite at the Chandrasekhar mass? – further cloud the issue of $^{60}$Fe production by Type Ia events. We estimate that these high ignition density models may increase the steady state injection rate of $^{60}$Fe by about 0.15 M$_\odot$ Myr$^{-1}$, although these models are probably not typical Type Ia supernova events.

## 4. The 1.8 MeV Map of Galactic $^{26}$Al emission

The longitude-latitude maps of Diehl *et al.* (1994, 1995) tend to show that the emission is generally confined in latitude to within 5 degrees of the Galactic plane, and 50° in longitude from the Galactic center. There are clumps, local hot spots, voids and distincts enhancement due to the vicinity of spiral arms. Figure 5a shows the latitude integrated ($\pm$ 5°), relative intensity longitude profile of COMPTEL as the solid curve. This curve reflects the limited longitudinal extent (and other properties) present in the two-dimensional flux maps.

Since the chemical evolution model assumed an infinitely thin disk, to calculate the flux densities at various longitudes the vertical stratification of $^{26}$Al must be parameterized. We chose a simple exponential scale length of 200 pc for the chemical evolution postprocessing. This value was chosen primarily because it is a reasonable OB star scale height above the plane; other values simply rescale the calculated flux map. The effects of spiral structures in the Galactic disk were investigated with a model that uses the free electron distribution of Taylor & Cordes (1993) as a tracer of $^{26}$Al. In this approach one sums over all the number density line of sight integrals of the radioactive species. The functional form of these integrals are similar to the dispersion measure integral that appears in studies of radio pulsars (see Hartmann *et al.* 1995 for details). Spiral arms and local features in the three-dimensional distribution give variations in the calculated flux map as a function of longitude and latitude and has been invoked by several studies of 1809 keV line emission from the Galactic plane (Chen *et al.* 1994; Prantzos 1994), A 200 pc vertical scale height for the simple chemical evolution model also yields fluxes that are comparable to the multi-dimensional model.

Figure 5b shows the latitude integrated flux profile that emerges from the chemical evolution radial profile (dashed curve) and the multi-dimensional model (solid curve). The calculated curves are symmetric about zero longitude and thus cannot explain the observed global asymmetries. However, the calculated flux maps show that the flux drops rapidly outside of longitudes within $\pm$ 30°, as indicated by the observation shown in Figure 5a . However, our models appear slightly too centrally condensed. The relatively flat radial source profile implied by the COMPTEL observations is incompatible with a steeply decreasing radial source distribution. In addition, there are hot spots. Some of the hot spot enhancements in our map, due to spiral arm features, coincide with peaks in the observed data but they are not as strong. It is presently not clear how much of the emission along the Galactic plane is due to discrete foreground sources, but some of the observed structure should be due to the global spiral arm structure. Thus, while the calculated flux distributions of Figure 5b display some general properties of the observed flux map in Figure 5a, it does not get all the details correct.

## 5. Sources of Uncertainty

Unlike elemental ratios such as [O/Fe], which are relatively insensitive to the chemical evolution model, absolute yields, such as the steady state abundances of $^{26}$Al and $^{60}$Fe, are much more dependent upon assumptions and uncertainties. This is especially true for isotopes that are trace constituents of the composition. The uncertainties in nuclear cross section, stellar evolution, chemical evolution and flux mappings all contribute to the error.

The nuclear reaction rates into and out of $^{26}$Al used in this survey are from Caughlan & Fowler (1988), with the exception of $^{25}$Mg(p,$\gamma$)$^{26}$Al which has been adjusted by us to include new information on low-lying resonances reported by Iliadus et al. (1990). Our expression agrees with their calculated stellar rate to within 20%, and differs from the Caughlan & Fowler (1988) result only at low temperatures ($T_9 \leq 0.25$), where the new rate is slightly smaller. Therefore, the $^{26}$Al contribution from hydrogen shell burning could have been larger. The explosive yields from the oxygen-neon shell, where the bulk of the $^{26}$Al is made at peak temperatures well above $T_9 = 0.9$, were unaffected by using the Iliadis rate.

The neutron capture reaction rates affecting $^{60}$Fe synthesis are from Woosley et al. (1978). The (n,$\gamma$) reactions on $^{56,57,58}$Fe, have been adjusted to agree with the experimentally measured 30 keV cross sections recommended by Bao & Kappeler (1986), [$\sigma_{n\gamma}$(30 keV) = 13.0, 35.2, and 13.0 mb, respectively, for $^{59,60}$Fe, $\sigma_{n\gamma}$(30 keV) = 12.0, and 3.59 mb, respectively]. For a complete list of reaction rate references, see Woosley & Hoffman (1990) and Hoffman & Woosley (1995).

The $^{12}$C($\alpha$,$\gamma$)$^{16}$O rate effects not only the composition of the ashes after core helium burning, but the entropy and density structure of the star as it continues to evolve. The rate we used was the Caughlan & Fowler (1988) value multiplied by 1.7. This is equivalent to S(300 keV) = 170 keV barns, a value that has been determined to be optimal for producing the solar abundance set (Weaver & Woosley 1993). This value is also supported by recent measurements; S(300 keV) = 79 ± 21 or 82 ± 26 keV barns (R- and K-matrix fits, respectively) for the E1 part of this rate (Azuma et al. 1994); the experimental and theoretical expectation that E2 is $\sim$ 70 ± 50 kev barns (Barnes 1995; Mohr et al. 1995). Other resonances also contribute to the S-factor at 300 keV at the level of approximately 19 KeV barns, so that the best current combined experimental and theoretical estimate is 169 ± 55 keV barns (Barnes 1995).

As a check on the Woosley & Weaver (1995) results, the solar metallicity 15 and 25 M$_\odot$ stars were evolved with a much larger nuclear reaction network (315 isotopes up to and including 14 isotopes of krypton). After recalculating the entire evolution (presupernova + explosion) the mass of $^{26}$Al and $^{60}$Fe ejected by the 15 M$_\odot$ model deviated from the Woosley & Weaver (1995) results by 0.9% and -1.3% respectively, while the changes for the 25 $M_\odot$ star were 1.8% and -1.9%. The error in the abundances of these isotopes due to the extent of network used by Woosley & Weaver (1995) is negligible.

Convection has been treated within the boundaries defined by mixing-length theories. Multidimensional treatments (Bazan & Arnett 1994) of convective oxygen burning (where both $^{26}$Al and $^{60}$Fe are produced) have shown that chemically distinct plumes of material rising and falling, vortices and local hot spot burning may significantly effect the yields of rare isotopes from single stars.

The total baryonic mass of the Galaxy determines how easy it is to make trace constituents; a larger reservoir requires less processing to produce a given amount of a contaminant. All other parameters being equal, it is twice as easy to make 2 $M_\odot$ of $^{26}$Al in a Galaxy that has a total baryonic (gas + stars) mass of $2 \times 10^{11}$ $M_\odot$ than it is with a Galaxy that has only $1 \times 10^{11}$ $M_\odot$, since the more massive galaxy will generally have a larger Type II + Ib supernova rate. How the total baryonic mass is distributed (e.g, the choice of the scale length of the exponential disk) affects the amount of $^{26}$Al and $^{60}$Fe produced. A smaller disk scale length concentrates more mass in the central regions, where larger star formation rates are generally found. The slope of the initial mass function, along with the integration limits, determine how many massive stars exist. The more negative the exponent in the Salpeter function, the steeper the initial mass function and the less massive stars there are to form the radioactive nuclides. The star formation rate, and hence the Galactic core collapse rate, is probably the most critical parameter. The amount of $^{26}$Al and $^{60}$Fe produced is linear with variations of the efficiency of star formation (or equivalently the Galactic supernovae rate). A core collapse rate one half as large produces one half the yields.

Approximately 100 chemical evolution models were calculated varying the total baryonic mass of the Galaxy ($6 \times 10^{10}$ $M_\odot \leq M_{Gal} \leq 3 \times 10^{11}$ $M_\odot$), the mass distribution ($1 \times 10^3$ $M_\odot \leq M_{center} \leq 2 \times 10^5$ $M_\odot$), the slope of the Salpeter initial mass function ($-1.7 \leq x \leq -1.2$), the upper mass limit of the initial mass function (25 $M_\odot \leq M_{up} \leq 100$ $M_\odot$), the efficiency of star formation ($0.8 \leq \nu \leq 10.8$) and the exponent in the Schmidt star formation rate ($1.0 \leq k \leq 2.0$). Each variation was performed independently and the injection rates of $^{26}$Al and $^{60}$Fe determined. The solar abundance pattern was examined in each of these models, but no injection rate was discarded based on the quality of the fit. The most sensitive parameters were found to be the total baryonic mass of the Galaxy, the rate of core collapse events and the upper mass limit of the initial mass function. Examination of the output from this sensitivity study gives the quoted error bars on the present epoch injection rate of $2.0 \pm 1.0$ $M_\odot$ Myr$^{-1}$ for $^{26}$Al and $0.75 \pm 0.4$ $M_\odot$ Myr$^{-1}$ for $^{60}$Fe. These error bars only take into account the uncertainties in the chemical evolution model.

There are uncertainties inherent in mapping the one-dimensional chemical evolution model into the multi-dimensional flux map model, primarily the choice of the vertical scale height and overall normalization. The total mass in the three-dimensional model of the present day Galaxy is uncertain, which effects the amplitude of the tracer population, and hence the absolute amplitude of the flux. The number of spiral arms could change the number of hot spot features attributable to spiral arms. Diehl *et al.* (1994, 1995) suggest that some particular features of the emission may be due to foreground sources. The authors estimate that only $\sim 1$ $M_\odot$ of $^{26}$Al may be needed to explain the underlying smooth emission distribution. It may be possible to adjust the Galactic chemical evolution parameters to simultaneously achieve a good fit to the solar abundances, produce a core collapse rate of 1.5 or even 1 per century, and still satisfy the 1809 keV observations.

## 5. Conclusions and Predictions

The production sites, spatial distribution and chemical evolution of the gamma-ray producing nuclei $^{26}$Al, and $^{60}$Fe were examined. Our model suggests that core collapse supernovae presently provide an $^{26}$Al injection rate of $2.0 \pm 1.0$ $M_\odot$ Myr$^{-1}$, and $0.75 \pm 0.4$ $M_\odot$ Myr$^{-1}$ of $^{60}$Fe, with the error bars reflecting variations in the assumptions and chemical evolution models. This corresponds to $2.2 \pm 1.1$ $M_\odot$ of $^{26}$Al and $1.7 \pm 0.9$ $M_\odot$ of $^{60}$Fe in the present interstellar medium. Sources of uncertainty were discussed, with one of the more important being the present rate of core collapse supernovae in the Galaxy. When these results were mapped into a multidimensional model of the present Galaxy, the calculated 1.8 MeV $^{26}$Al flux maps were shown to be consistent with the steep decline of the observed flux outside a longitude of $\pm 30°$, and the the slight enhancement in the vicinity of spiral arms.

When $^{60}$Fe is detected in the interstellar medium we predict that the flux maps will be concentrated toward the Galactic center, the $^{60}$Fe mass and flux distributions will follow the $^{26}$Al distributions (giving added weight to the assertion that Type II supernovae are responsible for most of the $^{26}$Al in the Galaxy), the $^{60}$Fe and $^{26}$Al hot spots will overlap, and $0.75 \pm 0.4$ $M_\odot$ of live $^{60}$Fe will be inferred to exist in the present epoch interstellar medium. If Chandrasekhar mass white dwarfs explode frequently from high central densities as Type Ia supernova, then the prediction for the steady state abundance of $^{60}$Fe increases by $\sim 0.15$ $M_\odot$ Myr$^{-1}$. The resulting mass and flux distributions of $^{60}$Fe would then have a systematic offset from the $^{26}$Al distributions.

This work has been supported at Clemson by NASA (NAG 5-1578); at Santa Cruz by the NSF (AST 91 15367), NASA (NAGW 2525) and the California Space Institute (CS86-92); at Livermore by the Department of Energy (W-7405-ENG-48) and at Chicago by an Enrico Fermi Postdoctoral Fellowship (FXT).

TABLE 1

Massive Star $^{26}$Al and $^{60}$Fe Yields[a]

| Mass | $^{26}$Al | $^{60}$Fe | Mass | $^{26}$Al | $^{60}$Fe |
|------|-----------|-----------|------|-----------|-----------|
| 11 | 1.68E-05 | 8.78E-06 | 20 | 3.47E-05 | 1.12E-05 |
| 12 | 2.00E-05 | 2.91E-06 | 22 | 5.91E-05 | 5.19E-05 |
| 13 | 2.84E-05 | 1.05E-04 | 25 | 1.27E-04 | 2.10E-05 |
| 15 | 4.30E-05 | 2.66E-05 | 30 | 2.73E-04 | 2.38E-05 |
| 18 | 8.14E-05 | 2.54E-05 | 35 | 3.47E-04 | 5.59E-05 |
| 19 | 8.83E-05 | 4.69E-05 | 40 | 2.51E-04 | 8.32E-05 |

[a] All entries in solar masses for solar metallicity stars.

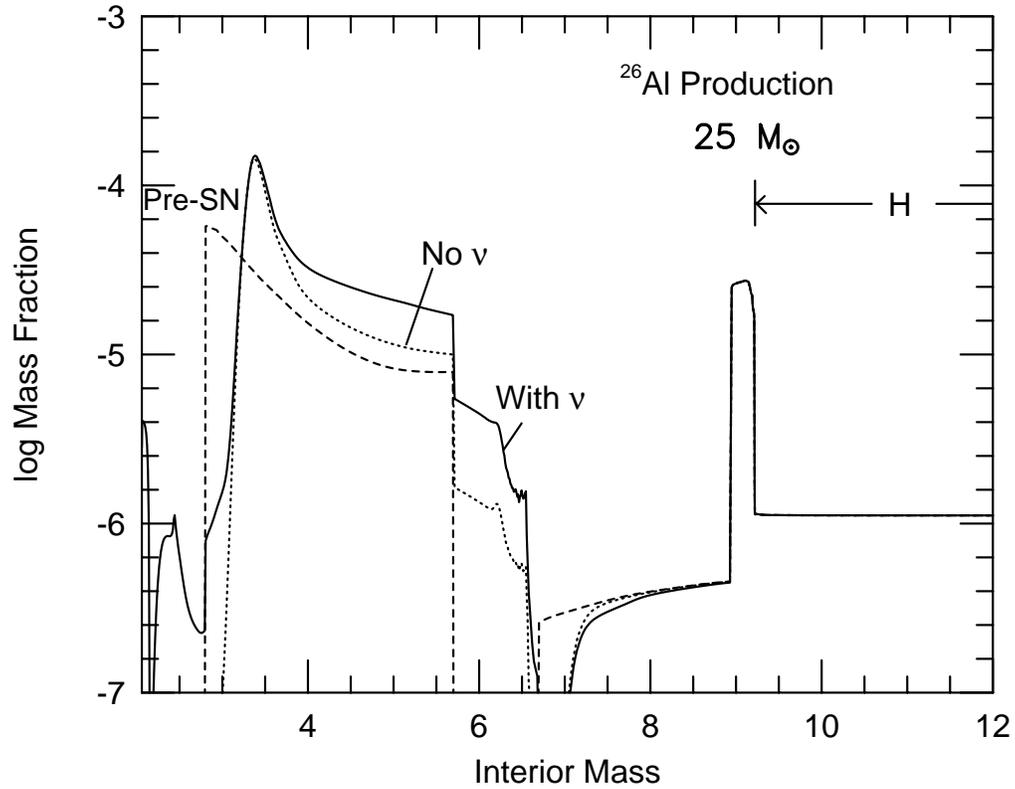

Fig. 1.— Mass fraction of $^{26}$Al vs interior mass (M$_\odot$) for a solar metallicity 25 M$_\odot$ star. Production in the presupernova star (dashed line), explosion (dotted line) and enhancement to the explosive processing by the $\nu$-process (solid line) are shown.

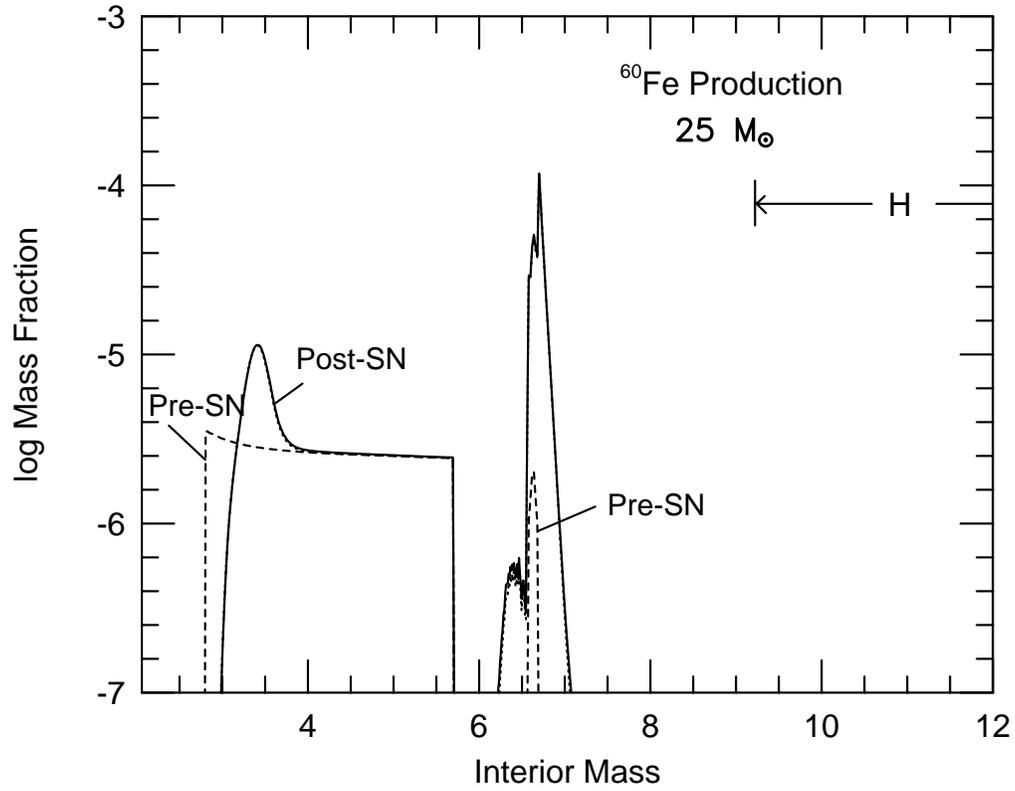

Fig. 2.— Mass Fraction of $^{60}$Fe vs interior mass ($M_\odot$) for a solar metallicity 25 $M_\odot$ star. Production in the presupernova star (dashed line), and explosion (solid line) are shown. The $\nu$-process did not contribute to the explosive yield.

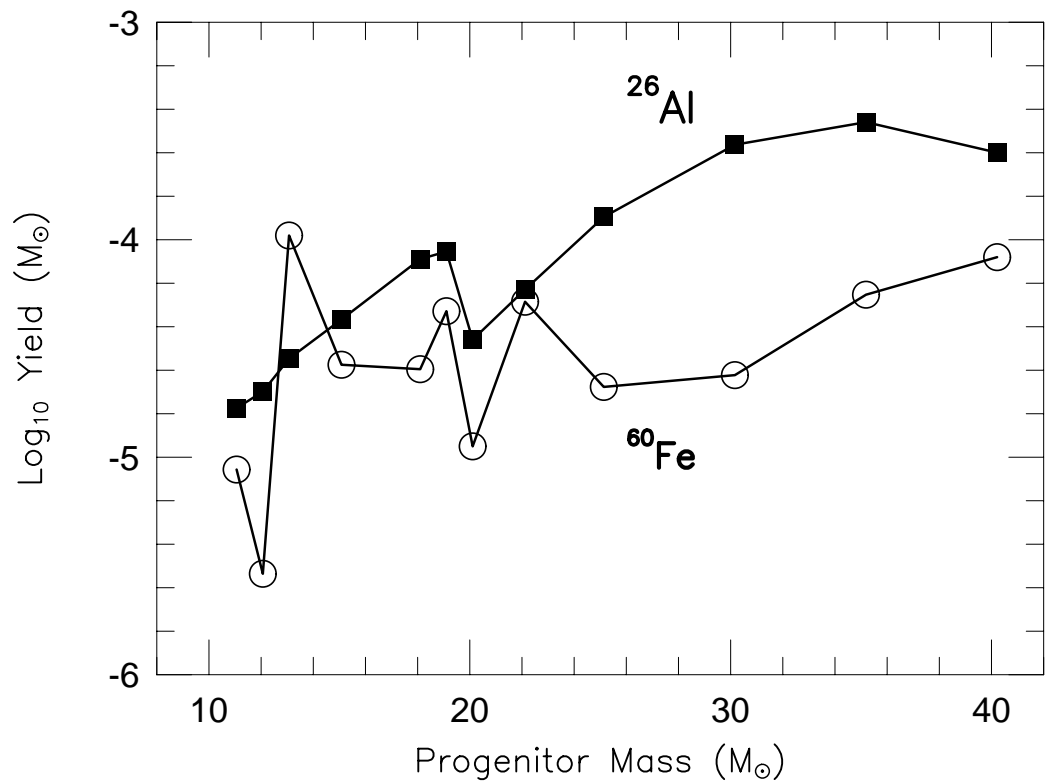

Fig. 3.— Mass of $^{26}$Al (filled squares) and $^{60}$Fe (open circles) ejected as a function of the main–sequence progenitor mass.

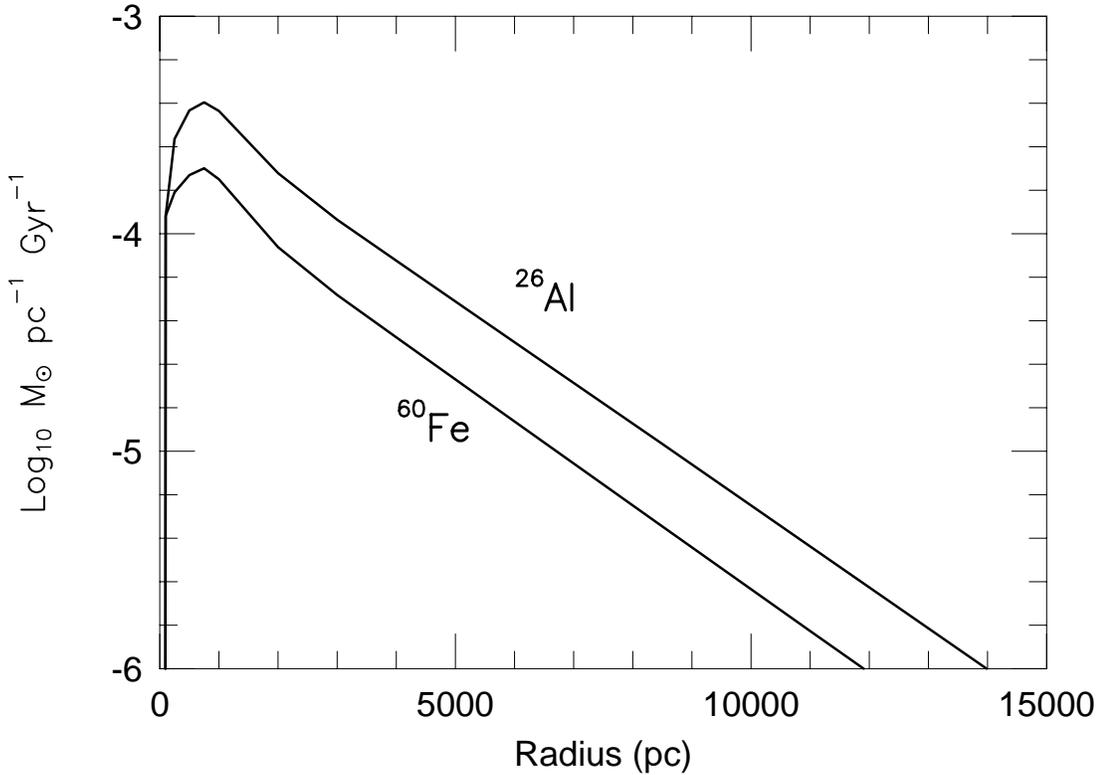

Fig. 4.— Injection rate (times $2\pi r$) of $^{26}$Al and $^{60}$Fe as a function of Galactocentric radius. These radial distributions are a reflection of the assumed exponential Galactic disk and should not be interpreted as *a priori* determinations. The figure shows that the distribution of these radionuclides will be strongly concentrated toward the center of the Galaxy. The area under the curve gives the injection rates. Including estimates of the uncertainties (see text), the integrations give present epoch injection rates of $2.0 \pm 1.0$ M$_\odot$ Myr$^{-1}$ for $^{26}$Al and $0.75 \pm 0.4$ M$_\odot$ Myr$^{-1}$ for $^{60}$Fe.

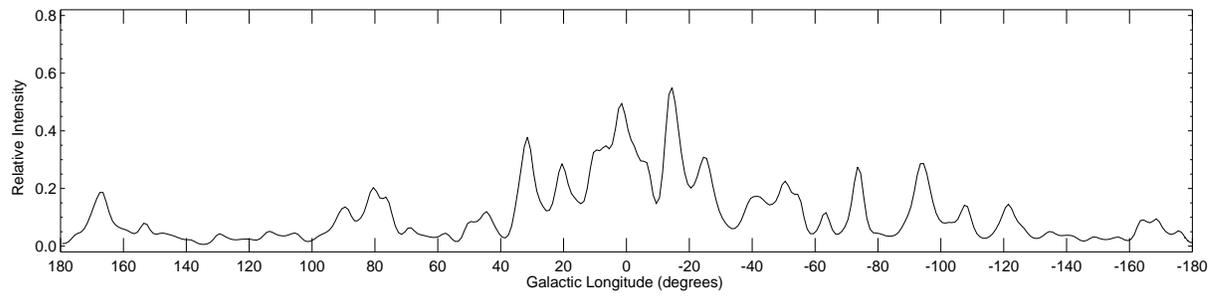

Fig. 5a.— Intensity of the 1809 KeV line from $^{26}$Al integrated between $\pm$ 5° as a function of Galactic longitude. The observations are from Diehl *et al.* (1994, 1995).

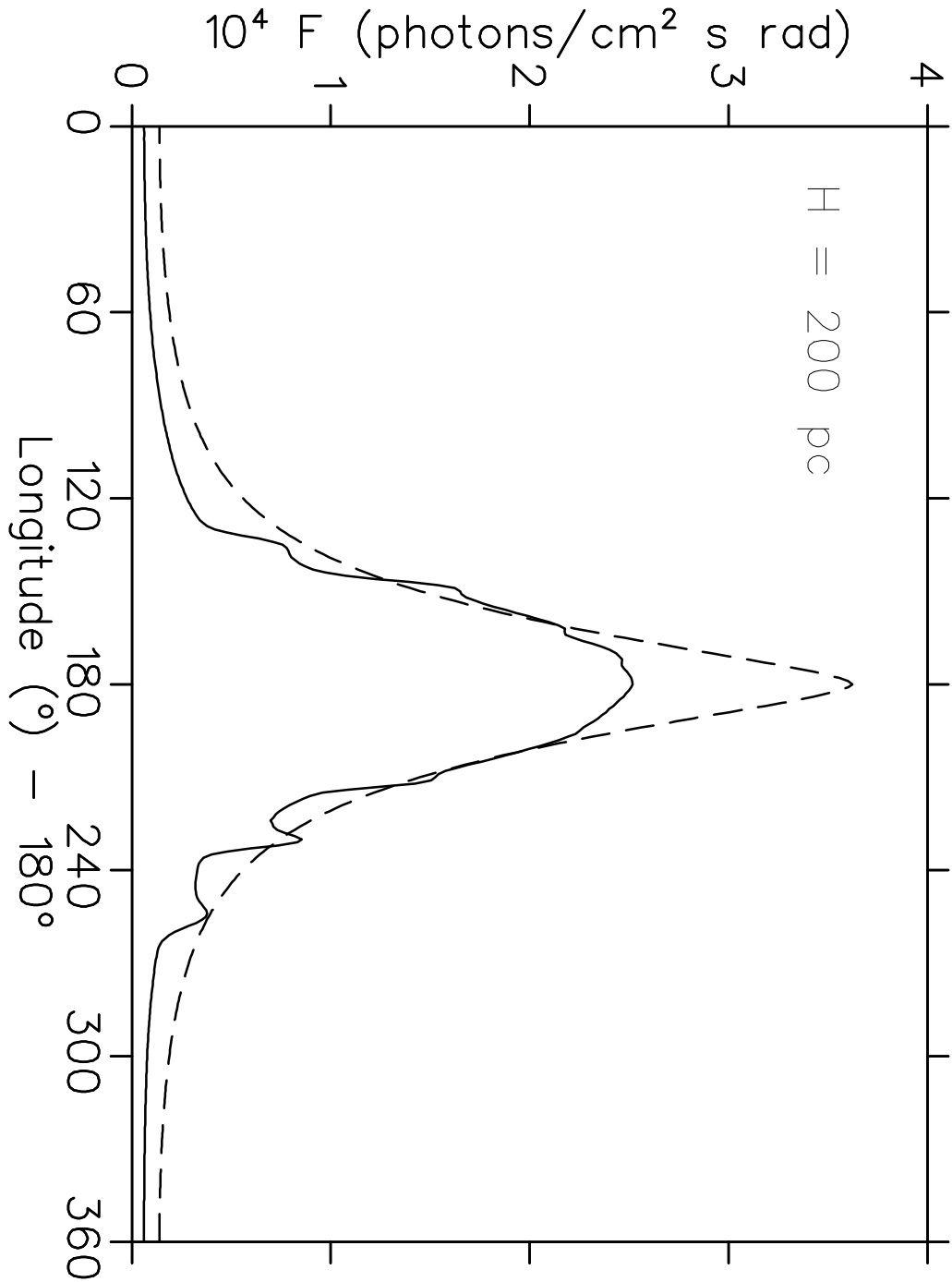

Fig. 5b.— Intensity of the 1809 KeV line from $^{26}$Al integrated between ± 5° as a function of Galactic longitude. The dashed curve is from the chemical evolution model, with the addition of a vertical scale height of 200 pc. The solid curve follows from the convolution of the chemical evolution model with the tracer model of Taylor & Cordes (1993) and exhibits spiral arm features.